\documentclass[12pt]{article}
\usepackage{paper2e}
\usepackage{mydefs2e}
\usepackage{epsf}

\newcommand{\beqa}{\begin{eqnarray}}
\newcommand{\eeqa}{\end{eqnarray}}
\newcommand{\msusy}{\ensuremath{\frac{F}{M}}}
\newcommand{\msusysquare}{\ensuremath{\bigg\vert \frac{F}{M}\bigg\vert^2}}

\def\caption#1{{\centerline{\vbox{\baselineskip=14pt
          \vskip.15in\hsize=5.5in\noindent{#1}\vskip.1in }}}}

\begin{document}
\begin{titlepage}

\preprint{LBNL-49279}
\preprint{YCTP-P6-01}

\begin{center}
\Large\bf
Yukawa Deflected Gauge Mediation 
\end{center}

\author{Z. Chacko$^{\bf a}$ \footnote{zchacko@thsrv.lbl.gov} and
\ Eduardo Pont\'{o}n$^{\bf b}$ \footnote{eduardo.ponton@yale.edu}}

\address{$^{\bf a}$ Department of Physics,
University of California, Berkeley, CA 94720, USA \\
Theoretical Physics Group, Lawrence Berkeley National Laboratory,  \\
Berkeley, CA 94720, USA
\medskip \\ $^{\bf b}$ Department of Physics, Yale University,
New Haven, CT 06511, USA}

\begin{abstract}
We consider models which are natural extensions of those where
supersymmetry is broken at low energy scales and transmitted to
visible matter by gauge interactions. We investigate the situation
where the quark and lepton superfields of the MSSM are localized to
a brane in a higher dimensional space while the messenger fields
and the sector which breaks supersymmetry dynamically are localized
to another brane in the same space. The MSSM gauge and Higgs fields
are assumed to propagate in the bulk. If some of the messenger
fields and the Higgs fields have the same quantum numbers, this
allows the possibility of mixing between these fields so that the
physical Higgs and messenger fields are admixtures of the brane and
bulk fields. This manifests itself in direct couplings of the quark
and lepton fields to the physical messengers that are proportional
to the MSSM Yukawa couplings and hence preserve the
flavor structure of the CKM matrix. The result is new
contributions to the soft supersymmetry breaking parameters that
are related to the Yukawa couplings and which therefore naturally
satisfy the constraints from FCNC's. For messenger scales greater
then 1000 TeV these new contributions are parametrically of the
same order of magnitude as gauge mediation. This scenario naturally
avoids the cosmological problems associated with stable messengers
and admits a simple and natural solution to the $\mu$ problem based on the
NMSSM.

\end{abstract}


\end{titlepage}

\section{Introduction}

Gauge mediated supersymmetry breaking is arguably the most
attractive candidate for a realistic mechanism of supersymmetry
breaking \cite{gaugemed}. In this scenario one assumes that there
is a hidden sector in which supersymmetry is broken, and which
couples to a set of messenger fields charged under the standard
model gauge interactions. Supersymmetry breaking effects are then
communicated to the visible sector fields through loop effects
involving the gauge interactions. This leads to a viable and highly
predictive spectrum of sparticles. Since supersymmetry breaking is
communicated by gauge interactions the squark and slepton spectrum
is nearly flavor diagonal and therefore in good agreement with the
experimental constraints on flavor changing neutral currents
(FCNCs).

In gauge mediation it is usually assumed that direct interactions
between the MSSM fields and the messenger fields, if any, are very
small since this would lead to new sources of flavor violation
beyond the CKM matrix \cite{Dine:1996xk,Han:1998xy}. The current
constraints on flavor changing neutral currents place tight
constraints on any such interactions \cite{Gabbiani:1996hi}:
\begin{equation}
\frac{{m^2}_{ds}}{{m^2}_{ss}} \le (6 \times 10^{-3}) \frac{m_{ss}}{1
{\mbox{TeV}}}~.
\end{equation}

In this paper we consider a natural extension of gauge mediation
with messenger Higgs mixing in which there are no sources of flavor
violation apart from the CKM matrix itself. We consider the situation
where the quark and lepton superfields of the MSSM are localized to
a brane in a higher dimensional space, while the gauge and Higgs
fields propagate in the bulk. The messenger fields and the
supersymmetry breaking sector are assumed to be localized to
another brane in the same space. If some of the messenger fields
have the same quantum numbers as the Higgs fields, this allows the
possibility of mixing between them so that the physical Higgs and
messenger fields are admixtures of the brane and bulk fields. This
manifests itself in the Lagrangian as direct couplings of the quark
and lepton fields to the physical messengers that are proportional
to the MSSM Yukawa couplings and therefore preserve the
flavor structure of the CKM matrix. The result is new
contributions to the soft scalar masses that are related to the
Yukawa couplings of the standard model fermions, and which
therefore naturally satisfy the constraints from FCNC's. The extra
dimensions are assumed to be sufficiently small that four
dimensional gauge coupling unification is unaffected.
This also
allows other potentially large sources of supersymmetry breaking
such as anomaly mediation \cite{RS0}, gaugino mediation \cite{gMSB} and
radion mediation \cite{RMSB}
to be neglected. This scenario, which we call `Yukawa Deflected
Gauge Mediation', naturally avoids
the cosmological problems associated with stable messengers. We
further investigate the $\mu$ problem of Yukawa Deflected Gauge
Mediation in the context of the NMSSM. In the context of a specific
model, we demonstrate that it is indeed possible to generate the
correct pattern of symmetry breaking with a realistic spectrum of
masses.

Our idea is in the spirit of an earlier suggestion by Dvali and
Shifman \cite{Dvali:1996hs} that the Higgs doublets of the MSSM are
in fact also the messengers of gauge mediated supersymmetry
breaking. In that case there are also contributions to the scalar
masses related to the Yukawa couplings and constrained by the CKM
matrix, but obtaining a light Higgs doublet is not simple.

\section{Messenger-Higgs Mixing}

Consider a gauge mediated SUSY breaking model with 2 pairs of messengers
$(Q_m,\bar{Q}_m)$ (m = 1, 2) transforming as ${\bf 5}$ and ${\bf
\bar{5}}$ of SU(5).  Under SU(3)$\times$SU(2)$\times$U$(1)_Y$ these
decompose as ${\bf (3,1,-\frac{2}{3})\oplus(1,2,-1)}$ and ${\bf
(\bar{3},1,\frac{2}{3})\oplus(1,2,1)}$.  The SU(2) doublets in the
messengers which we denote by $Q_{iu}$ and $\bar{Q}_{id}$ have the same
quantum numbers
as the MSSM Higgs fields ${\tilde{H}}_u$ and ${\tilde{H}}_d$ and can
therefore mix with them.  They can also have additional Yukawa
couplings to the MSSM quarks $q_i$ and $u^c_i$ and leptons $l_i$ and
$e^c_i$.  In general these new Yukawa couplings will lead to
additional flavor violation outside the CKM matrix, and must therefore
be forbidden by a symmetry, such as the messenger number symmetry
which exists if the only messenger coupling in the superpotential is
$XQ\bar{Q}$.  Here $\langle X \rangle = M + F \theta^2$ is a chiral
superfield that parameterizes supersymmetry breaking.

We now consider the situation where the MSSM quarks and leptons are
localized to a brane in a five dimensional space, while the MSSM
gauge and Higgs fields live in the bulk of the space.  The
messenger fields and the sector which breaks supersymmetry
dynamically \cite{dynSUSY} are assumed to live on another brane.
The extra dimension is assumed to be sufficiently small that $1/r
\ge {\rm{M}}_{\rm GUT}$ and gauge coupling unification goes through
exactly as in four dimensions. However $r$ is assumed to be
sufficiently larger than the inverse cutoff of the higher
dimensional theory so that the exchange of massive bulk states with
mass of order the cutoff does not alter our conclusions about the
form of the effective theory below the scale $1/r$. Here we assume
that there are no other light bulk fields beyond those of
supergravity and the MSSM gauge and Higgs fields.

A 5D gauge multiplet consists of the gauge field $A_M$ ($M = 0,
\ldots, 4$), a real adjoint scalar $\sigma$, and a fermion
$\lambda$. We assume that the $5^{\rm th}$ dimension is
compactified on a $S^1 / Z_2$ orbifold of radius $r$. The fixed
points of the orbifold are `branes' on which the hidden and visible
sectors can be localized. The $Z_2$ parity assignments of the gauge
field are such that $A_5$, $\sigma$, and half of the $\lambda$
components are odd. These states will then get masses of order
$1/r$, and the surviving degrees of freedom make up an $\scr{N} =
1$ gauge multiplet (see \eg \Refs{MP,AGW} for details.) A 5D
hypermultiplet consists of 2 N=1 chiral multiplets one of which is
necessarily even and the other odd under the orbifold. Once again
the odd states are projected out and are not present in the
effective theory below the scale $1/r$. Therefore the Higgs
doublets of the MSSM ${\tilde{H}}_u$ and ${\tilde{H}}_d$ are
assumed to emerge from two different hypermultiplets.

As a consequence of the higher dimensional nature of the theory any Yukawa
couplings between the messenger fields and the MSSM quarks and leptons are
forbidden by locality. However, mixing between the messengers and the
Higgs fields is still allowed. After integrating out the extra dimension
the superpotential of the higher dimensional theory has the form
\beq\eql{effectiveL} \bal \!\!\!\!\!\! W = & X \left[\sum_{m=1}^2
\lambda_m Q_{m}{\bar{Q}}_{m} + \tilde{\lambda}_d Q_{1u} {\tilde{H}}_d +
\tilde{\lambda}_u {\tilde{H}}_u {\bar{Q}}_{2d} \right] + \\ & \hspace{4mm}
 \left[{\tilde{y}}_{U,ij} {\tilde{H}}_u q_i u^c_j +
{\tilde{y}}_{D,ij}{\tilde{H}}_d q_i d^c_j +
{\tilde{y}}_{L,ij}{\tilde{H}}_d l_i e^c_j \right] +
\\
& \hspace{4mm}
\left[\mu {\tilde{H}}_u {\tilde{H}}_d + {\mbox{Gauge Kinetic Terms}}\right].
\eal\eeq
where in order to avoid the dangerous term
$X{\tilde{H}}_u{\tilde{H}}_d$ we have imposed the discrete symmetry $
X \rightarrow - X, Q_{1u} \rightarrow - Q_{1u}, {\bar{Q}}_{2d}
\rightarrow -{\bar{Q}}_{2d} $, with all other fields neutral.

From this expression it is clear that the physical doublet messengers are
\begin{eqnarray}
{{\bar M}}_{1d} &=& \frac{\tilde{\lambda}_d {\tilde{H}}_d +
\lambda_1{\bar{Q}}_{1d}}{\sqrt{
{\tilde{\lambda}_d}^2 + {\lambda_1}^2}} \\ M_{2u} &=&
\frac{\tilde{\lambda}_u {\tilde{H}}_u + \lambda_2{{Q}}_{2u}}{\sqrt{
{\tilde{\lambda}_u}^2 + {\lambda_2}^2}}~,
\end{eqnarray}
while the physical Higgs fields $H_u$ and $H_d$ are the orthogonal linear
combinations. The superpotential rewritten in terms of these fields takes the
form
\beq
\label{rewrittenL}
\bal
\!\!\!\!\!\!
W = &X \left[\sum_{m=1}^2 \lambda_m Q_{mT}{\bar{Q}}_{mT} +
\lambda_1'Q_{1u}{{\bar M}}_{1d} + \lambda_2'M_{2u}{{\bar Q}}_{2d}\right] + \\
& \hspace{4mm} \left[ y_{U,ij} {{H}}_u q_i u^c_j + y_{D,ij}{{H}}_d
q_i d^c_j + y_{L,ij}{H}_d l_i e^c_j \right] +
\\& \hspace{4mm}
 \left[ y_{U,ij}' M_{2u} q_i u^c_j + y_{D,ij}'{{\bar M}}_{1d} q_i d^c_j + y_{L,ij}'{{\bar M}}_{1d}
l_i e^c_j \right] + \cdots
\eal\eeq
where $Q_{mT}$ and ${\bar{Q}}_{mT}$ denote the messenger
$SU(3)_{C}$ triplets.  The new couplings $\lambda', y$ and $y'$ are
related in a straightforward way to the old couplings $\lambda$ and
$\tilde{y}$. In particular, note that the ratios
\begin{eqnarray}
\label{proportionality}
\frac{y_{U,ij}}{y_{U,ij}'} &=& {{k_U}} \\
\frac{y_{D,ij}}{y_{D,ij}'} &=& {{k_D}}
\end {eqnarray}
are independent of the indices $i$ and $j$. This implies that the Yukawa
couplings of the messengers to matter are proportional to the MSSM Yukawa
couplings. Therefore the new supersymmetry breaking effects that emerge
from the direct messenger matter couplings will be constrained by the CKM
matrix and the sizes of the Yukawa couplings and will not give rise to
large flavor violation.

Since the messenger doublets now have direct renormalizable
couplings to the visible sector fields they are no longer stable
and can directly decay into them. However one may worry that this
is not true of the the messenger triplets and that these will be
stable leading to cosmological difficulties. However, if the theory
emerges from a supersymmetric grand unified theory the messenger
triplets mix with the Higgs triplets, which have direct couplings
to matter. The Higgs triplets are integrated out at the GUT scale.
Then in the effective theory below the Higgs triplet mass there are
direct couplings of the messenger triplets to visible fields
suppressed by powers of the Higgs triplet mass. While these
couplings are renormalizable and dimensionless they are small, of
order $M/ {\rm M}_{\rm GUT}$. Nevertheless they are easily large
enough to allow the triplets to decay sufficiently rapidly so as to
avoid the cosmological problems associated with stable messengers.

We now attempt to determine the size of the supersymmetry breaking
contributions from these new direct messenger-matter interactions.
The one loop contributions to the scalar mass$^2$ from the Yukawa
type couplings to the messengers vanish to leading order in
$(F/M)^2$ \cite{Dine:1996xk}. The subleading one loop contributions
of order $[y'^2/(16\pi^2)] (F^4/M^6)$ are smaller than the leading
two loop contributions which are of order
$[y'^2/(16\pi^2)][g_3^2/(16\pi^2)](F/M)^2$ provided $(F/M^2) \le
g_3/(4\pi)$. Notice that the two loop contributions are always
parametrically of the same order as the usual gauge mediated
contributions. Hence we will concentrate on the case where the
messenger scale is large, $M \gg 10^6$ GeV, when the one loop
contributions to the scalar masses can be safely neglected. Other
contributions to the soft terms are trilinear A terms which arise
at one loop.

In the next section we give a derivation of the most general two
loop contributions to the soft masses and the one loop
contributions to the A terms, at the messenger scale. Below we give
the expressions for the model Eq.~(\ref{rewrittenL}) keeping only
the Yukawa couplings for third generation particles.

With the notation $y_{t} \equiv y_{U,33}$, $y_{b} \equiv y_{D,33}$,
$y_{\tau} \equiv y_{L,33}$ and similarly for the new, primed Yukawa
couplings, we find the following expressions for the new
contributions to the soft masses at the messenger scale:
\beqa
\Delta m^2_{\tilde q_3} &=& \frac{1}{128 \pi^4}\left[ {y'_{b}}^2
\left( - \frac{8}{3} g_3^2 - \frac{3}{2} g_2^2 - \frac{14}{9} g_Y^2 +
{y'_{t}}^{2} + 3 {y'_{b}}^2 + \frac{1}{2} {y'_{\tau}}^2 \right)
\right. \nonumber \\ & & \hspace{5.3cm} \mbox{} + {y'_{t}}^{2} \left.
\left( - \frac{8}{3} g_3^{2} - \frac{3}{2} g_2^{2} - \frac{26}{9} g_Y^{2} +
3 {y'_{t}}^{2} \right) \right] \msusysquare \nonumber \\
\Delta m^2_{\tilde t^c} &=& \frac{1}{128 \pi^4} \left[ {y'_{t}}^{2}
\left( - \frac{16}{3} g_3^{2} - 3 g_2^{2} - \frac{52}{9} g_Y^{2} +
6 {y'_{t}}^{2} + y_{b}^{2} + {y'_{b}}^2 \right) - {y'_{b}}^2
y_{t}^2 \right]
\msusysquare
\nonumber \\
\Delta m^2_{\tilde b^c} &=& \frac{1}{128 \pi^4} \left[ {y'_{b}}^{2}
\left( - \frac{16}{3} g_3^{2} - 3 g_2^{2} - \frac{28}{9} g_Y^{2} +
y_{t}^{2} + {y'_{t}}^2 + 6 {y'_{b}}^{2} + {y'_{\tau}}^{2} \right) -
{y'_{t}}^2 y_{b}^2 \right]
\msusysquare \nonumber \\
\Delta m^2_{\tilde l_3} &=& \frac{{y'_{\tau}}^2}{128 \pi^4}
\left( - \frac{3}{2} g_2^2 - 6 g_Y^2 + \frac{3}{2} {y'_{b}}^2 +
2 {y'_{\tau}}^2 \right)
\msusysquare \\
\Delta m^2_{\tilde \tau^c} &=& 2 \Delta m^2_{\tilde L_i} \nonumber \\
\Delta m^2_{H_u} &=& - \frac{3 y_{t}^2}{256 \pi^4} \left[
6 {y'_{t}}^2 + {y'_{b}}^2 \right] \msusysquare
\nonumber \\
\Delta m^2_{H_d} &=& - \frac{1}{256 \pi^4} \left[
3 y_{b}^2 \left({y'_{t}}^2 + 3{y'_{b}}^2 \right) + 3
y_{\tau}^{2}{y'_{\tau}}^{2} +
\left(3 y_{b} y'_{b} + y_{\tau} y'_{\tau} \right)^{2} \right]
\msusysquare~. \nonumber
\eeqa
Here $g_Y$ is the hypercharge gauge coupling where the hypercharge
is defined by $Q = T_3 + Y/2$, and $T_3$ is the third SU(2)
generator. These have to be added to the well known gauge mediated
expressions
\beqa
\label{gaugemediation}
m^2_{\tilde q_3} &=& \frac{N}{128 \pi^4}
\left(\frac{20}{27}g_Y^4 + \frac{3}{4}g_2^4 + \frac{4}{3}g_3^4\right) \nonumber \\
m^2_{\tilde t^c} &=& \frac{N}{128 \pi^4}
\left(\frac{320}{27}g_Y^4 + \frac{4}{3}g_3^4\right) \nonumber \\
m^2_{\tilde b^c} &=& \frac{N}{128 \pi^4}
\left(\frac{80}{27}g_Y^4 + \frac{4}{3}g_3^4\right) \nonumber \\
m^2_{\tilde l_3} &=& \frac{N}{128 \pi^4}
\left(\frac{20}{3}g_Y^4 + \frac{3}{4}g_2^4\right) \\
m^2_{\tilde \tau^c} &=& \frac{N}{128 \pi^4}
\left(\frac{80}{3}g_Y^4\right) \nonumber \\
m^2_{Hu,Hd} &=& \frac{N}{128 \pi^4}
\left(\frac{20}{3}g_Y^4 + \frac{3}{4}g_2^4\right)~, \nonumber
\eeqa
where $N$ is the number of ${\bf 5} \oplus {\bf \bar{5}}$ messenger
pairs. We also find the following one loop contributions to the
A--terms:
\beqa
A_{t} &=& \frac{y_{t}}{16 \pi^2} \left( 3 {y'_{t}}^2 + {y'_{b}}^2
\right) \msusy \nonumber \\
A_{b} &=& \frac{y_{b} }{16 \pi^2} \left({y'_{t}}^2 + 3 {y'_{b}}^2
\right) \msusy \\ A_{\tau} &=&
\frac{3 y_{\tau} {y'_{\tau}}^2}{16 \pi^2} \msusy~.  \nonumber
\eeqa

The expressions above show that the new contributions to the scalar masses
are comparable to those from gauge mediation for the up sector of the
third generation. This is also true for the down sector if tan$\beta$ is
large. Even for tan$\beta \approx 10$ and $y_b \approx y_b'$ Yukawa
deflection gives a $10\%$ correction to the mass$^2$ of the right handed
sbottom at the messenger scale.

\section{Derivation of the Soft Terms}

In this section we derive the general expressions for the soft
\susy breaking terms induced at the messenger scale.  These results
can then be applied to theories with matter-messenger couplings
like the models we are considering.  The general formulae are most
easily derived by the method of analytical continuation into
superspace developed in \cite{Giudice:1997ni}. We start by
reminding the reader of the basic idea.  If \susy breaking in the
messenger sector is parameterized by the VEV of a chiral superfield
$\langle X \rangle = M + F \theta^{2}$, then the leading \susy
breaking contribution to the observable sector, in an expansion in
powers of $F/M^2$, can be described within a \susc framework.  More
precisely, if the parameters of the theory at a scale
$\Lambda_{UV}$ above the messenger scale $M$ are fixed, then the
low-energy values of the wave function renormalization constants
will depend, through their RG evolution, on the scale $M$ at which
the messenger fields are integrated out.  The soft \susy breaking
parameters can then be incorporated by the replacement $M
\rightarrow |X|$ in the \Kahler potential, that is by analytical
continuation into superspace (in holomorphic terms the correct
analytical continuation is given by $M \rightarrow X$).  If the
observable sector superfields are denoted by $Q'^{a}$, the
low-energy Lagrangian, in the presence of soft \susy breaking, can
then be written as
\beq
\label{GeneralLagrangian}
\scr{L} = \int d^{4}\theta Q'^{\dagger}_{a} \scr{Z}(|X|)^{a}_{\;b} Q'^{b} +
\left( \int d^{2}\theta \lambda'_{abc} Q'^{a} Q'^{b} Q'^{c} + h.c. \right)~.
\eeq
For simplicity, here we chose to show only the Yukawa couplings in
the superpotential. The generalization to other operators will be
evident in what follows. Note also that we allow for off-diagonal
mixing in the kinetic terms, so that $\scr{Z}(|X|)$ is a general
hermitian matrix. The soft \susy breaking terms can be read from
the Lagrangian (\ref{GeneralLagrangian}) after replacing $X$ by its
VEV and expanding in powers of $\theta$: $\scr{Z}(|X|) = Z +
\frac{1}{2}
\frac{\partial Z}{\partial M} (F
\theta^{2} + F^{\dagger} \bar\theta^{2}) + \frac{1}{4}
\frac{\partial^{2} Z}{\partial M^{2}} F F^{\dagger} \theta^{2}
\bar\theta^{2}$, where $Z = \scr{Z}(M)$ is the usual wave function
renormalization constant.  To display these terms more clearly, it
is convenient to perform the following (chiral) field redefinition
$Q = Z^{1/2} \left(1 + Z^{-1} \frac{\partial Z}{\partial M} F
\theta^{2} \right) Q'$, after which the Lagrangian becomes
\beqa
\label{RescaledLagrangian}
\scr{L} &=& \int d^{4}\theta Q^{\dagger}_{a} Q^{a} +
\left( \int d^{2}\theta \lambda_{abc} Q^{a} Q^{b} Q^{c} + h.c.
\right) \nonumber \\
& & \hspace{2cm} \mbox{} - \tilde Q^{\dagger}_{a}
\left({m_{\tilde Q}^{2}}\right)^{a}_{\;b} \tilde Q^{b} -
\left(A_{abc} \tilde Q^{a} \tilde Q^{b} \tilde Q^{c} + h.c. \right)~.
\eeqa
Here $\tilde Q$ is the scalar component of Q, $\lambda_{abc} =
\lambda'_{a'b'c'} (Z^{-1/2})^{a'}_{\;a}
(Z^{-1/2})^{b'}_{\;b} (Z^{-1/2})^{c'}_{\;c}$ are the renormalized
Yukawa couplings, and the soft masses are given by
\beq
\label{softmassesGeneral}
m_{\tilde Q}^{2} = - \frac{1}{4} Z^{-1/2} \left(
\frac{\partial^{2} Z}{\partial \ln M^{2}} - \frac{\partial Z}{\partial \ln M}
Z^{-1} \frac{\partial Z}{\partial \ln M} \right) Z^{-1/2}
\frac{F F^{\dagger}}{M M^{\dagger}}
\eeq
while the A--terms are given by
\beqa
\label{AtermsGeneral}
A_{abc} &=& \frac{1}{2} \left(
\lambda_{a'bc} \left[ Z^{-1/2} \frac{\partial Z}{\partial \ln M}
Z^{-1/2}\right]^{a'}_{\;\;a} +
\lambda_{ab'c} \left[ Z^{-1/2} \frac{\partial Z}{\partial \ln M}
Z^{-1/2}\right]^{b'}_{\;\;b} \right. \nonumber \\ & & \left.
\hspace{2.5cm} \mbox{} +
\lambda_{abc'} \left[ Z^{-1/2} \frac{\partial Z}{\partial \ln M}
Z^{-1/2}\right]^{c'}_{\;\;c} \right) \msusy~.
\eeqa
In order to find explicit expressions for the soft parameters
(\ref{softmassesGeneral}) and (\ref{AtermsGeneral}) at a scale
$\mu$ one needs to solve for $Z(\mu;M)$ from its RG evolution
equation
\beq
\label{RGforZ}
\frac{d Z}{dt} = \gamma Z~,
\eeq
where $\gamma$ is the matrix of anomalous dimensions and $t=\ln
\mu$. In general, it is not possible to find closed expressions for
the soft \susy breaking parameters even at lowest loop order,
except in a few simple cases \cite{Giudice:1997ni}.  It is however
possible to write closed expressions for the soft parameters at the
scale $\mu = M$, which can then be used as initial data for a
numerical solution to the RG equations below the messenger scale.
This is the strategy that we will follow and our next task is to
find the general formulae for the soft parameters just below the
messenger scale. From Eqs.~(\ref{softmassesGeneral}) and
(\ref{AtermsGeneral}), we see that we need to evaluate the first
and second derivatives of $Z(\mu;M)$ with respect to $\ln M$. In
order to do this, we first note that by rescaling the fields, we
can conveniently set $Z(\Lambda_{UV}) = 1$. Writing then $Z = 1 +
\delta Z$ at an arbitrary scale and integrating Eq.~(\ref{RGforZ}),
we formally obtain for scales $\mu < M$
\beq
\delta Z(t;M) = \int_{\ln \Lambda_{UV}}^{\ln M} dt' \gamma_{>}(t') [ 1 +
\delta Z(t')] + \int_{\ln M}^{t} dt' \gamma_{<}(t';M) [ 1 +
\delta Z(t';M)]~. \nonumber
\eeq
In writing this expression we took into account the fact that the
anomalous dimensions can be discontinous at $\mu = M$, and denoted
by $\gamma_{>}$ ($\gamma_{<}$) the anomalous dimensions above
(below) $M$.  Our notation also reflects the fact that the
anomalous dimensions as well as $\delta Z$ can depend on $M$ only
\textit{below} the messenger scale.  Differentiating once with respect to
$\ln M$ we find
\beqa
\label{FirstDerivative}
\frac{d \delta Z(t;M)}{d \ln M} &=&
\Delta \gamma(M) [1 + \delta Z(M)] + \\
& & \int_{\ln M}^{t} dt'
\left\{\frac{d \gamma_{<}(t';M)}{d \ln M} [ 1 + \delta Z(t';M)] +
\gamma_{<}(t';M) \frac{d \delta Z(t';M)}{d \ln M} \right\}~, \nonumber
\eeqa
where $\Delta \gamma(M) \equiv \gamma_{>}(M) - \gamma_{<}(M)$ and
we defined $\gamma_{<}(M) \equiv \gamma_{<}(t\!=\!\ln M; M)$. To
obtain Eqn.~(\ref{FirstDerivative}) we also used the fact that
$\delta Z$ is continuous across $M$ (the anomalous dimensions are
finite) so that $\delta Z(M) = \int_{\ln \Lambda_{UV}}^{\ln M} dt'
\gamma_{>}(t') [ 1 + \delta Z(t')]$ is well defined.  Taking now a
second derivative and evaluating at $\mu = M$ we obtain
\beqa
\frac{d^{2} \delta Z(t;M)}{d \ln M^{2}} \bigg\vert_{t = \ln M} &=&
\frac{d \Delta \gamma(M)}{d \ln M} [1 + \delta Z(M)] +
\Delta \gamma(M) \frac{d \delta Z(M)}{d \ln M} \nonumber \\
& & \mbox{} - \frac{d \gamma_{<}(t;M)}{d \ln M} [ 1 + \delta Z(M)] -
\gamma_{<}(M) \frac{d \delta Z(t;M)}{d \ln M} \bigg\vert_{t = \ln M}~.
\nonumber
\eeqa
In order to simplify this expression, we note that (in a
mass-independent scheme) the anomalous dimensions depend on $\ln M$
only through the gauge and Yukawa couplings, {\bf both} of which
will be generically denoted by $\lambda$. This implies that
$\frac{d \gamma}{d \ln M}$ is of 2-loop order and the terms
proportional to $\delta Z(M)$ are 3-loop effects, which we will
neglect. Using also $\frac{d \delta Z(M)}{d \ln M} = \gamma_{>}(M)$
and, from Eq.~(\ref{FirstDerivative}), $\frac{d \delta Z(t;M)}{d
\ln M}|_{t =
\ln M} \simeq \Delta \gamma(M)$, we can write
\beqa
\frac{d^{2} \delta Z(t;M)}{d \ln M^{2}} \bigg\vert_{t = \ln M} &=&
\sum_\lambda \frac{d \Delta \gamma(M)}{d \lambda(M)} \frac{d \lambda(M)}{d \ln M} +
\Delta \gamma(M) \gamma_{>}(M) - \gamma_{<}(M) \Delta \gamma(M) \nonumber \\
& & \mbox{} - \sum_\lambda \frac{d \gamma_{<}(t;M)}{d \lambda(t;M)}
\frac{d \lambda(t;M)}{d \ln M} \bigg\vert_{t = \ln M} +
\mathrm{(3\!-\!loop\;order)} \nonumber
\eeqa
It only remains to evaluate $\frac{d \lambda(t;M)}{d \ln M}$, which
we can do starting from the corresponding RG equation. If
$\beta[\lambda]$ is the $\beta$-function for $\lambda$, we can
formally write for $\mu < M$
\beq
\lambda(t;M) = \lambda(\Lambda_{UV}) +
\int_{\ln \Lambda_{UV}}^{\ln M} dt' \beta_{>}[\lambda(t')]
+ \int_{\ln M}^{t} dt' \beta_{<}[\lambda(t';M)]~. \nonumber
\eeq
Differentiating with respect to $\ln M$ and evaluating at $\mu =
M$, we get
\beq
\frac{d \lambda(t;M)}{d \ln M}\bigg\vert_{t = \ln M} =
\Delta \beta[\lambda(M)]~,
\eeq
where $\Delta \beta[\lambda(M)] \equiv
\beta_{>}[\lambda(M)]-\beta_{<}[\lambda(M)]$ and $\lambda(M) =
\lambda(\Lambda_{UV}) + \int_{\ln \Lambda_{UV}}^{\ln M} dt'
\beta_{>}[\lambda(t';M)]$. From the expression for $\lambda(M)$
we also see that $\frac{d \lambda(M)}{d \ln M} =
\beta_{>}[\lambda(M)]$. The second derivative can then be put in
the following form:
\beqa
\label{SecondDerivative}
\frac{d^{2} \delta Z(t;M)}{d \ln M^{2}} \bigg\vert_{t = \ln M} &=&
\sum_\lambda \left(\frac{d \Delta \gamma(M)}{d \lambda(M)} \beta_{>}[\lambda(M)] -
\frac{d \gamma_{<}(M)}{d \lambda(M)} \Delta \beta[\lambda(M)] \right) \nonumber \\
& & \hspace{1cm} \mbox{} + (\Delta \gamma(M))^2 + [\gamma_{>}(M),
\gamma_{<}(M)]~,
\eeqa
where $[A,B] = A B - B A$ is a commutator. We have now all the
ingredients required to evaluate the soft parameters at the
messenger scale. To lowest loop order we can replace all factors of
$Z$ by 1 in Eqs.~(\ref{softmassesGeneral}) and
(\ref{AtermsGeneral}). Then using Eqs.~(\ref{FirstDerivative}) and
(\ref{SecondDerivative}) (evaluated at $\mu = M$), we obtain the
final 2-loop expressions for the soft masses. In matrix notation
these are
\beqa
\label{softmasses}
m_{\tilde Q}^{2} \big\vert_{\mu = M} = - \frac{1}{4} \left\{
\sum_\lambda \left(\frac{d \Delta \gamma}{d \lambda} \beta_{>}[\lambda] -
\frac{d \gamma_{<}}{d \lambda} \Delta \beta[\lambda] \right) +
[\gamma_{>}, \gamma_{<}]  \right\} \bigg\vert_{\mu = M}
\frac{F F^{\dagger}}{M M^{\dagger}}~.
\eeqa
For the A--terms, we obtain from Eqs.~(\ref{AtermsGeneral}) and
(\ref{FirstDerivative}) the 1--loop result
\beq
\label{Aterms}
A_{abc} \big\vert_{\mu = M} = \frac{1}{2} \left(
\lambda_{a'bc} \Delta \gamma^{a'}_{\;\;a} +
\lambda_{ab'c} \Delta \gamma^{b'}_{\;\;b} +
\lambda_{abc'} \Delta \gamma^{c'}_{\;\;c} \right) \bigg\vert_{\mu = M} \msusy~.
\eeq
Equations (\ref{softmasses}) and (\ref{Aterms}) are the main
results of this section. These equations are understood to hold
just below the messenger scale. In particular, the sums in
Eq.~(\ref{Aterms})) run only over the couplings in the effective
low-energy theory. Given a specific model it is now straightforward
to calculate the induced soft terms at the messenger scale. Note
that in the absence of direct matter-messenger couplings the
anomalous dimensions of the observable fields are continuous at
$\mu = M$. In this case only the second (and third) terms in Eqn.
(\ref{softmasses}) survive and one recovers the standard gauge
mediated results when $\lambda$ is a gauge coupling.

\section{The $\mu$ problem}

The models of Yukawa Deflected Gauge Mediation naturally satisfy
all constraints coming from neutral flavor changing processes,
which could arguably be considered the most difficult challenge in
theories of \susy breaking.  A second issue that should be
addressed in any model of \susy breaking is the origin of the Higgs
bilinear term in the superpotential
\cite{Giudice:1988yz,Dvali:1996cu,Yanagida:1997yf,Dimopoulos:1997je,
Langacker:1999hs,Mafi:2000kg,Babu:2001gp}
\beq
\label{muterm}
W = \mu H_u H_d~.
\eeq
In its most basic form the difficulty arises because, for
phenomenological reasons, $\mu$ should be of the order of the weak
scale.  This scale is in turn related to the scale of \susy
breaking (if the hierarchy problem is to be solved by \susy) and
there is \textit{a priori} no reason that the
\susc term (\ref{muterm}) should be of weak scale order.  It is
then natural to assume that the $\mu$-term vanishes at tree-level
and is generated only after \susy breaking, for example from
\Kahler terms like \cite{Giudice:1988yz}
\beq
\label{Kahlermu}
K = \lambda H_u H_d \left( \frac{X^\dagger}{M} +
\frac{X X^\dagger}{M^2} + \cdots \right)~,
\eeq
where $\langle X \rangle = M + F \theta^2$. After \susy breaking
the first term in (\ref{Kahlermu}) generates the $\mu$ term
(\ref{muterm}) while the second generates the \susy breaking term
\beq
\label{Bmuterm}
V = B \mu H_u H_d~.
\eeq

As has been stressed in \cite{Dvali:1996cu}, in theories of gauge
mediation the real challenge is to explain why $B$ and $\mu$ are of
the same order. Since all other soft masses are generated at one
loop, one needs $\lambda \sim 1/16 \pi^2$ in order that $\mu \sim
(1/16 \pi^2) F/M$ has the correct size.  The problem is then that
$B \mu \sim (1/16 \pi^2) (F/M)^2$ which implies the relation $B
\sim (16 \pi^2) \mu$. Indeed, generically both the $\mu$ and $B
\mu$ terms are generated at the same loop order, which results in
the previous relation.  Such a large value of $B$ would require an
unacceptable degree of fine-tuning to obtain a correct electroweak
symmetry breaking pattern.

A very appealing solution to this problem is to introduce a new
light standard model singlet field $S$ with superpotential
couplings \cite{Agashe:1997kn,deGouvea:1997cx,Han:1999jc}
\beq
\label{WNMSSM}
W = \lambda S H_d H_u - \frac{1}{3} \kappa S^3~.
\eeq
If \susy breaking gives a negative mass-squared to $S$, then in the
process of electroweak symmetry breaking it will acquire a VEV and
an effective $\mu = \lambda \langle S \rangle$ of the correct size
will be generated.  Similarly, the $B \mu$ term can arise from the
A--term
\beq
\label{BmuNMSSM}
V = A_\lambda S H_u H_d~.
\eeq
Unfortunately, in models of gauge mediation, both $m_s^2$ and
$A_\lambda$ are very small and it has been shown that it is not
possible to obtain a realistic symmetry breaking pattern
\cite{deGouvea:1997cx}. In addition, there is always a light state
associated with the spontaneous breaking of an approximate
R-symmetry under which all superfields have R-charge $2/3$. This
symmetry is only broken by the term Eq.~(\ref{BmuNMSSM}), which is
very small in gauge mediation.

On the other hand, in models of Yukawa deflected mediation A--terms
are generated at one loop as we have shown in Eq.~(\ref{Aterms})
and thus they have the required order of magnitude to destroy the R
symmetry.
  Also, in these models $m_S^2$ can get a substantial negative
contribution, which can lead to a sizable $S$ VEV. In this section
we analyze the Next-to-Minimal \Susc Standard Model (NMSSM),
defined by the replacement of the $\mu$-term in the MSSM by the
superpotential Eq.~(\ref{WNMSSM}), and show that it is possible to
obtain realistic electroweak symmetry breaking.

We pause to note that in this model there is no \susc CP problem
\cite{Dugan:1984qf}. By redefining the phases of $S$ and $H_u$, we
can assume without loss of generality that $\lambda$ and $\kappa$
are real. By rescaling ${{\bar M}}_{1d}$ and $M_{2u}$ in
Eq.~(\ref{rewrittenL}) we can assume that the proportionality
constants $k_U$ and $k_D$ in Eq.~(\ref{proportionality}) are real
as well.  We can also assume that the couplings $\lambda_i$,
$\lambda'_i$ in the hidden brane are real by rotating the remaining
messenger fields.  Now all CP phases will reside in the CKM matrix.
In order to see this, one can rotate the matter superfields to the
quark mass eigenbasis.  By redefining the quark superfield phases,
one can absorb, as usual, all but one of the CKM phases.  In the
quark mass eigenbasis the gauge symmetry is not explicit and, in
particular, the Yukawa interactions between the matter and charged
Higgs superfields are not flavor diagonal whereas those involving
the neutral Higgses, by definition, are.  However, the important
point is that all Yukawa interactions can be written in terms of
the physical CKM matrix and the real quark mass eigenvalues.
Furthermore, since all field redefinitions are performed at the
superfield level, there are no additional phases in any of the soft
parameters.  Therefore there is only one physical CP violating
phase.

In what follows we neglect for simplicity the CP phase and assume that all
parameters are real.  We have the option of either restricting S to a
brane or allowing it to propagate in the bulk.  Allowing S to propagate in
the bulk allows for a greater range of couplings, since it can now couple
directly to the messenger triplets, as well as to the doublets.

In order to see the main features more easily, we will consider the
case in which only $\tilde H_d$ and $S$ propagate in the bulk.
Further we will neglect the terms involving the smaller Yukawa
couplings $y_b$ and $y_\tau$, as well as $y'_b$ and $y'_\tau$ which
are proportional to them.  If $\tan \beta$ is large, however, one
should also include these couplings. The superpotential we consider
at the messenger scale has the form
\beqa
W &=& y_{t} H_u q_3 t^c - S H_u (\lambda H_d + \lambda_d \bar M_{1d})
- \frac{\kappa}{3} S^{3}~.
\eeqa
While other couplings of S are in principle allowed by symmetry, we
are neglecting them here for purposes of simplicity. If such
additional couplings are not large we do not expect them to
significantly alter our conclusions. It is now straightforward to
obtain the soft breaking terms that are induced after integrating
out the messenger fields, from the general equations
(\ref{softmasses}) and (\ref{Aterms}).  We find in addition to the
standard gauge mediated contribution Eq.~(\ref{gaugemediation}),
the following nonvanishing new contributions to the soft masses of
the observable fields, at the messenger scale:
\beqa
\label{NMSSMmasses}
\Delta m^2_{\tilde q_3} &=& - \frac{\lambda_d^2 y_{t}^2}{256 \pi^4}
\msusysquare \nonumber \\
\Delta m^2_{\tilde t^c} &=& - \frac{\lambda_d^2 y_{t}^2}{128 \pi^4} \msusysquare \nonumber \\
\Delta m^2_{H_u} &=& \frac{\lambda_d^2}{128 \pi^4}
\left( 2 \lambda_d^2 + \kappa^2 - 2 g_Y^2 - \frac{3}{2} g_2^2 \right) \msusysquare \\
\Delta m^2_{H_d} &=& - \frac{\lambda_{d}^{2} \lambda^{2}}{64 \pi^4} \msusysquare \nonumber \\
\Delta m^2_{\tilde S} &=& - \frac{\lambda_d^2}{128 \pi^4}
\left(4 g_Y^2 + 3 g_2^2 + 4 \kappa^2 - 4 \lambda_d^2 -
3 y_{t}^2 \right) \msusysquare \nonumber
\eeqa
We observe that $m_S^2$ can indeed be negative if $\kappa
\sim 1$. In this region of parameter space $m^2_{H_u}$ receives also
a positive contribution. Similarly, the nonvanishing one-loop
trilinear terms are:
\beqa
\label{NMSSMAterms}
A_{t} &=& \frac{y_{t}\lambda_d^2}{16 \pi^2} \msusy \nonumber \\
A_{\lambda} &=& \frac{3\lambda \lambda_{d}^2}{16 \pi^2} \msusy \\
A_{\kappa} &=& \frac{\kappa \lambda_d^{2}}{8 \pi^2}
\msusy
\nonumber
\eeqa
These A--terms are defined in Eq.~(\ref{RescaledLagrangian}). In
particular, we have not factored out the corresponding Yukawa
coupling.

Eqs.~(\ref{NMSSMmasses}) and (\ref{NMSSMAterms}) are all
proportional to $\lambda_{d}$.  Given the values of the various
Yukawa couplings at the messenger scale as well as the \susy
breaking scale $F/M$, one can use the NMSSM renormalization group
equations, to obtain the values of the various soft masses at the
weak scale.  As usual, the Higgs mass parameter is driven negative
by the top Yukawa coupling and we find electroweak symmetry
breaking minima for a large range of parameters.  In order to
reproduce the $Z$ boson mass, $M_Z$, we require that the Higgs
VEV's satisfy $v^2 \equiv v_u^2 + v_d^2 = (174 \GeV)^2$.  This
fixes the overall scale.  The minimization also determines $\tan
\beta = \frac{v_u}{v_d}$, and one should try to adjust $y_t$ to
reproduce $m_{\rm{top}} \sim 165 \GeV$ (the difference with the
experimental value of about $175 \GeV$ is attributed to QCD
corrections.)  However, the fact that $y_t$ is attracted to its
low-energy quasi-fixed point, leaves some freedom in the choice of
$y_t$ at the messenger scale.  This choice is however important in
determining the evolution of various quantities such as
$m_{H_u}^2$.  In practice we take as arbitrary input parameters the
values of the Yukawa couplings $y_t, \lambda,
\kappa$ and $\lambda_d$ at the messenger scale as well as the
messenger scale $M_{\rm mess}$.

We give two sample points in Table~1. We used $g_Y = 0.1816$, $g_2
= 0.6486$ and $g_2 = 1.1005$ for the gauge coupling constants at
the 1 TeV scale, and checked that the theory remains perturbative
up to the GUT scale. The rest of the input parameters are given in
the table, as well as the weak scale values for the various
physical masses, which include the soft as well as the D-term
contributions.

We note that the NLSP is the right-handed stau (as in gauge
mediation, the gravitino is the LSP.) This is due to the effect of
the $U(1)_Y$ Fayet--Iliopoulos D--term \cite{D-term} in the RG
running of the soft masses,
\beq
\Delta \frac{d}{d t} m_i^2 = \frac{1}{16 \pi^2} Y_i g_Y^2 \sum_j Y_j m_j^2~,
\eeq
where the sum runs over all fields and $Y_i$ is the hypercharge of the
i-th field.  In pure gauge mediation this contribution vanishes, but
it is in general not zero in the presence of Yukawa couplings.  In our
case we find at the messenger scale
\beq
\sum_j m_j^2 = \frac{\lambda_d^2}{32 \pi^2} \left( 3 y_t^2 +
4 \lambda^2 + 4 \lambda_d^2 + 2 \kappa^2 - 4 g_Y^2 - 3 g_2^2\right)
\eeq
which is always positive since $y_t \sim 1$ cancels the smaller
negative gauge contributions. Therefore, the fields having a
positive (negative) hypercharge will receive a negative (positive)
contribution from this term. The most important effect is on the
right-handed sleptons and thus we expect the NLSP to correspond to
the stau in this class of models.

A second distinctive feature is the relation $\Delta m^2_{\tilde
t^c} = 2 \Delta m^2_{\tilde q_3}$, which holds, up to small
corrections proportional to $y_b, y_\tau$, even when both $H_u$ and
$H_d$ are allowed to propagate in the bulk.

When analyzing the spectrum at the weak scale it is important to
include the radiative corrections to the lightest neutral Higgs
mass \cite{radHiggs}. The largest effect can be viewed as a
top-stop loop contribution to an effective quartic term in the
effective potential below the stop mass \cite{Haber:1993an}. We
include an estimate of this effect by adding the term
\beq
\De V_H = \left( \frac{3 y_t^4}{8\pi^2}
\ln\frac{m_{\tilde{t}}}{m_t} \right)
(H_u^\dagger H^{\vphantom\dagger}_u)^2
\eeq
to the Higgs potential.

An important feature of these results is the amount of fine-tuning
required to achieve electroweak symmetry breaking. We define the
fractional sensitivity to a parameter $c$ (a coupling renormalized
at $M_{\rm Mess}$) to be \cite{Barbieri:1987fn,Anderson:1994dz}
\beq
\hbox{sensitivity} = \frac{c}{v} \frac{\partial v}{\partial c},
\eeq
where $v$ is the Higgs VEV and the derivative is taken with all
other couplings at the messenger scale held fixed. We find that the
largest sensitivities are associated with $\alpha_3 = g_3^2/(4
\pi)$ and $\lambda_d$. We note however that the sensitivities shown
in the table are of the same order as the ones one would obtain for
pure gauge mediation with tree level $\mu$ and $B_{\mu}$ terms
fixed by the requirement of correct electroweak symmetry breaking
(for the same values of $\tan \beta$ as shown in Table 1.) This
amount of fine-tuning seems to be inherent to models in which the
dominant soft breaking contributions arise from gauge mediation.

\begin{table}
\centering
\begin{tabular}{c|c|c|c}
&  & $N = 1$ & $N = 2$\\
\hline
inputs: & $M_{mess}$ & $10^{11}$ & $10^{14}$ \\ & $y_t$ & 0.9 & 0.9
\\ & $\lambda$ & 0.15 & 0.2 \\ & $\kappa$ & 0.8 & 0.98
\\ & $\lambda_d$ & 0.748 & 0.994 \\
\hline
neutralinos: & $m_{\chi^0_1}$ & 108 & 132 \\ & $m_{\chi^0_2}$ & 165
& 179
\\ & $m_{\chi^0_3}$ & 173 & 208 \\ & $m_{\chi^0_4}$ & 315 & 380
\\ & $m_{\chi^0_5}$ & 1550 & 1410 \\
\hline
charginos: & $m_{\chi^\pm_1}$ & 135 & 155 \\ & $m_{\chi^\pm_2}$ &
315 & 382 \\
\hline
Higgs: & $\tan\be$ & 7.1 & 5.5 \\ & $m_{h^0}$ & 115 & 115 \\ &
$m_{H^0}$ & 467 & 500 \\ & $m_A$ & 466 & 500 \\ & $m_{{H'}^0}$ &
1220 & 1170
\\ & $m_{A'}$ & 1660 & 1360 \\ & $m_{H^\pm}$ & 473 & 505 \\
\hline
sleptons: & $m_{\tilde{e}_R}$ & 100 & 113 \\ & $m_{\tilde{e}_L}$ &
470 & 505 \\ & $m_{\tilde{\nu}_L}$ & 465 & 500 \\
\hline
stops: & $m_{\tilde{t}_1}$ & 390 & 467 \\ & $m_{\tilde{t}_2}$ & 850
& 907
\\
\hline
other squarks: & $m_{\tilde{u}_L}$ & 1050 & 1105 \\ &
$m_{\tilde{u}_R}$ & 1000 & 1065 \\ & $m_{\tilde{d}_L}$ & 1050 &
1105
\\ & $m_{\tilde{d}_R}$ & 960 & 1010 \\
\hline
gluino: & $M_3$ & 815 & 1030 \\
\hline
sensitivity: & $\alpha_3$ & 100 & 130 \\ & $\lambda_d$ & 90 & 130
\\
\end{tabular}
\capt{\textbf{Table 1}. Sample points in parameter space for $N = 1, 2$ where N
is the number of ${\bf 5} \oplus {\bf \bar{5}}$ messenger pairs.
All masses are in GeV. $\alpha_3$ is the strong coupling constant
and the sensitivity parameter is defined in the main text.}
\end{table}

\section{Conclusions}

Yukawa deflection alters the spectrum of gauge mediated supersymmetry
breaking in a highly predictive manner while maintaining the requisite
suppression of flavor changing neutral currents. It is an important effect
for the third generation sparticles which have sizable Yukawa couplings.
We have demonstrated that it can resolve in a simple and natural way the
$\mu$ problem of gauge mediation, as well as the cosmological problems
associated with stable messengers.

\section{Acknowledgements}

We would like to thank Kaustubh Agashe, Emmanuel Katz, Markus A. Luty, Ann
E. Nelson, Elena Perazzi and Raman Sundrum for useful conversations at
various stages of this work. We would also like to thank the Aspen Center
for Physics for its hospitality. Z.C. was supported in part by the U.S.
Department of Energy under contract DE-ACO3-76SF00098 and in part by the
N.S.F. under grant PHY-00-98840. E.P. was supported by the DOE under
contract DE-FG02-92ER-40704.

\newpage


\end{document}